\documentclass[11pt,a4paper]{article}
\usepackage{amsmath,amscd}
\usepackage{amssymb}
\usepackage{cite}
\usepackage{comment}
\usepackage{graphicx}
\usepackage{hyperref}
\usepackage{mathpazo}
\usepackage{mathtools}
\usepackage{multimedia}
\usepackage{slashed}
\usepackage[small,nohug,heads=vee]{diagrams}
\usepackage[usenames,dvipsnames]{xcolor}
\usepackage{xfrac}

\definecolor{light-gray}{gray}{0.80}

\diagramstyle[labelstyle=\scriptstyle]

\newcommand\eq{=&&\hspace{-18pt}}
\newcommand\p{\hspace{1pt}}
\newcommand\strt[1]{\rule[-#1pt]{0pt}{#1pt}}

\begin{document}

\title{Mass as a dynamical quantity}
\author{M. Land \\ Hadassah College Jerusalem \\ martin@hac.ac.il}
\maketitle

\begin{abstract}
The Standard Model (SM) ascribes the observed mass of elementary particles to an
effective interaction between basis states defined without mass terms and 
a scalar potential associated with the Higgs boson.
In the relativistic field theory that underlies the SM, mass itself,
understood as the Lorentz-invariant squared 4-momentum of a particle or field, is
fixed \textit{a priori}, imposing a constraint on possible momentum states. 
Stueckelberg introduced an alternative approach, positing 
antiparticles as particles evolving backward in time, thus relaxing the mass
shell constraint for individual particles. 
Further work by Piron and Horwitz established a covariant Hamiltonian mechanics
on an unconstrained 8D phase space, leading to a gauge field theory that
mediates the exchange of mass between particles, while the total mass of
particles and fields remains conserved.   
In a recently developed extension of general relativity, consistent with this
approach, the spacetime metric evolves in a manner that permits the exchange of
mass across spacetime through the gravitational field. 
Mechanisms that restrict mass exchange between particles have also been
identified. 
Nevertheless, mass exchange remains possible under certain circumstances and may
have phenomenological implications in particle physics and cosmology. 
%
%
\end{abstract}


\section{Introduction}
\label{s:intro}
The Standard Model (SM) of elementary particles is a locally gauge invariant 
relativistic quantum field theory with particular choices for the basis states and
the gauge group.
While efforts to move beyond the SM usually begin by generalizing
the algebraic structure of the gauge fields, other work has focused on the
underlying framework of relativistic dynamics. 
In this paper, we present such an approach pioneered by Stueckelberg in 1941 
in his work on antiparticles. 
In this approach, particle mass is treated as a dynamical quantity,
leading to gauge theories in which fields and particles
may exchange mass, just as they exchange energy-momentum. 
We review the resulting classical and quantum electrodynamics, indicating mechanisms that maintain
particle masses at their familiar on-shell values.
Recent work on general relativity and gravitation is also outlined.  

The notion of an elementary particle characterized by a fixed mass can be traced
back to the 1897 discovery by Thomson \cite{Thomson} that cathode rays are composed
of discrete bodies with a fixed charge-to-mass ratio, and the 1909 Millikan-Fletcher
\cite{Millikan} oil-drop experiment indicating a minimum electron charge.
Today, the measured mass uncertainty of an electron is on the order of $\Delta m \simeq$
$10^{-8}$ \cite{PDG} and so it is conventional to write single-electron
equations of the type
\begin{equation}
\textcolor{black}{m} \frac{\hspace{1pt}\hspace{1pt} du^\mu}{d\tau} = e
F^{\mu\nu} u_\nu \qquad \qquad \left( i \slashed \partial - e\slashed A - 
\textcolor{black}{m} \right)\psi =0  
\label{p:3}
\end{equation}
for fixed $m$ and metric signature 
\begin{equation}
\eta_{\mu\nu} = \left( -,+,+,+ \right) \ .
\label{p:4}
\end{equation}
The fixed mass shell $ p^\mu p_\mu = - m^2 c^2$ is expressed in scattering by
writing
\begin{equation}
d^4p \hspace{2pt} \delta \big( p^\mu p_\mu +\textcolor{black}{m}^2 c^2
\big) = \frac{d^3{\mathbf{p}}}{2\sqrt{{\mathbf{p}}^2+
\textcolor{black}{m}^2 c^2}} 
\label{p:5}
\end{equation}
for the momentum space measure.
This picture is, of course, complicated by the SM, which defines a
Lagrangian containing no mass terms for elementary states, finding
effective mass terms through interaction with the symmetry-broken Higgs boson.  
Intriguingly, the effective masses of the composite nucleons $p$ and $n$ are
sharper ($\Delta m\sim 10^{-10}$) than the masses of their
constituent $u$ and $d$ quarks ($\Delta m\sim 25\% $) \cite{PDG}.  
The assumption of fixed masses is associated with a number of
issues in physics, including flavor oscillations, the problem of time, and
missing mass/energy in cosmology \cite{shi2021force}. 

\section{The Stueckelberg-Horwitz-Piron (SHP) framework}
\label{s:SHP}

A different approach was proposed by Stueckelberg \cite{Stueckelberg-1}
in his work on antiparticles.
Pair creation and annihilation are described in QED by the Feynman diagram 
in Figure 1a, showing an intermediate electron state propagating backward in time
with $E<0$, but observed as a positron with $E>0$ propagating forward in
time.
In the quantum picture, the electron jumps from forward timelike propagation to
backward timelike propagation, remaining on its mass shell throughout. 

But Stueckelberg proposed this model of pair processes in
{\em classical} electrodynamics, for a continuously evolving spacetime
trajectory $x^\mu(\tau)$ as shown in Figure 1b.
In such a curve, $\dot x^0 (\tau) = dx^0 / d\tau $ must vanish for some $\tau$,
and so $\dot x^\mu(\tau)$ must cross the spacelike region twice.
\begin{center}
\includegraphics[width=0.30\textwidth]{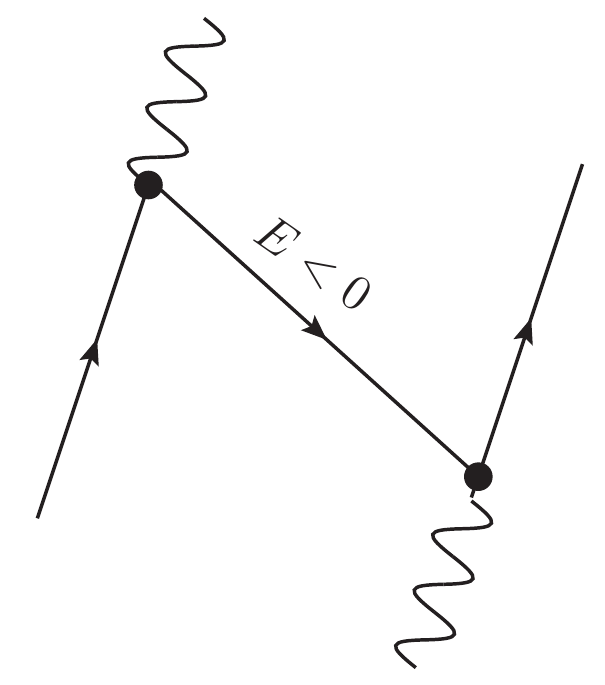} 
\qquad \qquad \qquad \qquad 
\includegraphics[width=0.25\textwidth]{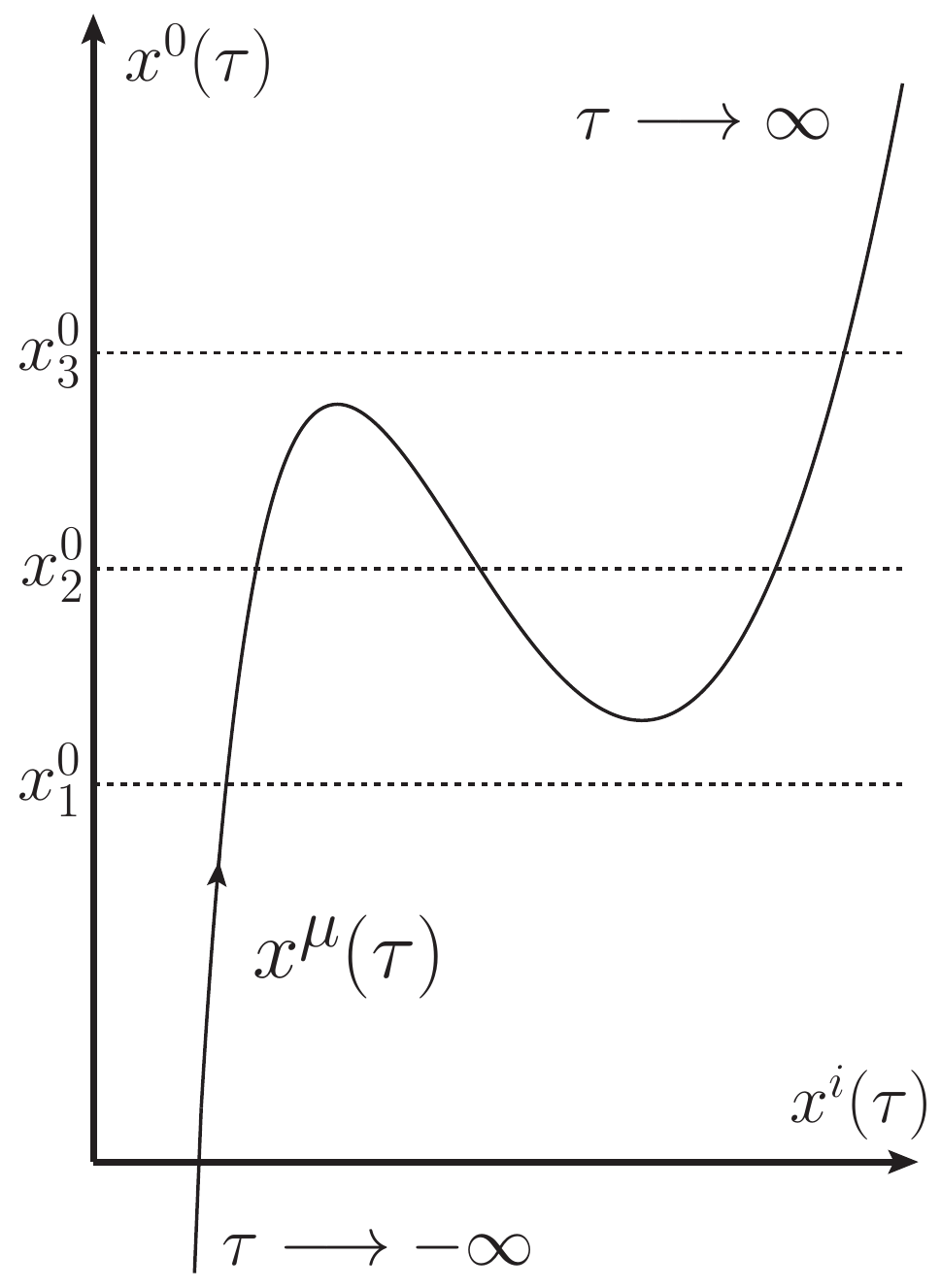} \\
\textbf{Figure 1a} \qquad \qquad \qquad \qquad\qquad \qquad \qquad \qquad
\textbf{Figure 1b} 
\end{center}
Stueckelberg observed that $\dot x^2 (\tau) = \dot x^\mu \dot x_\mu $ is a
dynamical quantity for this trajectory, and so all eight components of $\
x^\mu(\tau)$ and $ \dot x^\mu(\tau) $ must be independent. 
Moreover, since $\dot x^2 $ changes sign, the evolution cannot be 
parameterized by the proper time of the motion $ds = \sqrt{-\dot x^2 } d\tau $. 
The evolution parameter $\tau$ must be external to the spacetime
manifold, much as Newtonian time $t$ is external to space. 

Horwitz and Piron \cite{HP} were led to a similar model when constructing a
covariant canonical mechanics with non-trivial interactions.  
Writing a classical Lagrangian system on 8D unconstrained phase space
\begin{equation}
L=\frac{1}{2}M\dot{x}^{\mu }\dot{x}_{\mu }+e\dot{x}^{\mu }A_{\mu }\left(
x\right) - V \mbox{\qquad\qquad} \dfrac{d}{d\tau }\dfrac{\partial L}{
\partial \dot{x}_{\mu }}-\dfrac{\partial L}{\partial x_{\mu }}=0 
\label{Lagrangian}
\end{equation}
one obtains the generalized Lorentz force
\begin{equation}
M\left(\ddot{x}^{\mu }+\Gamma _{\nu \rho }^{\mu }\dot{x}^{\nu } \dot{x}%
^{\rho}\right) 
= eF^{\mu \nu }\dot{x}_{\nu } -\partial^\mu V 
\end{equation}
with field strength and conjugate momentum
\begin{equation}
F^{\mu \nu }=\partial ^{\mu }A^{\nu }-\partial ^{\nu }A^{\mu } %
\mbox{\qquad\qquad} p_\mu = \dfrac{\partial L}{\partial \dot{x}^{\mu }} = M%
\dot{x}_{\mu } + eA_{\mu }\left( x\right) \ .
\end{equation}
Transforming to the manifestly covariant Hamiltonian mechanics
\begin{equation}
K = \dot{x}^{\mu}p_\mu - L = \frac{1}{2M}(p^{\mu }-eA^{\mu })(p_{\mu
}-eA_{\mu }) +V 
\end{equation}
leads to the classical and quantum equations of motion
\begin{equation}
\dot{x}^{\mu }=\dfrac{\partial K}{\partial p_{\mu }} 
\qquad \dot{p}^{\mu }=-\dfrac{ \partial K}{\partial x_{\mu }}
\qquad \qquad i\partial _{\tau }\psi (x,\tau )= K\psi (x,\tau ) \ .
%
%
\end{equation}
Generalizing the classical central force problem as
\begin{equation}
V(x_1 , x_2) = V(\rho) \qquad \qquad \rho = \sqrt{(\mathbf{x}_1
- \mathbf{x}_2)^2 - (t_1 -t_2)^2}
%
%
\end{equation}
Horwitz and Arshansky \cite{bound-1}
obtained solutions for relativistic scattering and bound states, Horwitz and
Land studied radiative transitions, selection rules, perturbation theory, Zeeman and 
Stark effects, and bound state decay \cite{radiative-1},
and Horwitz demonstrated entanglement and interference in time
\cite{SHP-entanglement}.

The physical picture \cite{RCM} that emerges from Stueckelberg's unconstrained mechanics
can be summarized as an upgrade of nonrelativistic classical and quantum
mechanics in which Galilean symmetry is replaced with Poincar\'{e} symmetry:
\begin{equation*}
\left. 
\begin{array}{c}
\text{Newtonian time } t \vspace{00pt} \\ 
+ \vspace{-4pt} \\ 
\text{Unconstrained~} \left\{x^i,\dfrac{dx^j}{dt}\right\} \vspace{-2pt} \\ 
+ \vspace{00pt} \\ 
\text{Scalar Hamiltonian $H$ } \vspace{00pt} \\ 
\end{array}
\right\} \mbox{\quad}\longrightarrow \mbox{\quad} \left\{ 
\begin{array}{c}
\text{External time } \tau\vspace{00pt} \\ 
+ \vspace{-4pt} \\ 
\text{Unconstrained~} \left\{x^\mu,\dfrac{dx^\nu}{d\tau}\right\} \vspace{-2pt%
} \\ 
+ \vspace{00pt} \\ 
\text{Scalar Hamiltonian $K$ } \vspace{00pt} \\ 
\end{array}
\right. 
\end{equation*}
This covariant canonical mechanics inherits many methods and insights of
Newtonian mechanics, so for example, from the Poisson bracket
\begin{equation}
\{F,G\} = \frac{\partial F}{\partial x^\mu}\frac{\partial G}{\partial p_\mu}
- \frac{\partial G}{\partial x^\mu}\frac{\partial F}{\partial p_\mu}
\qquad \qquad 
\frac{dF}{d\tau} = \{F,K\} + \frac{\partial F}{\partial \tau}
%
%
\end{equation}
it follows that
\begin{equation*}
\dfrac{\partial H}{\partial t} = 0 \Rightarrow \text{conserved energy} \quad
\longrightarrow \quad \dfrac{\partial K}{\partial \tau} = 0 \Rightarrow 
\text{conserved mass} 
\end{equation*}
which for a free particle can be seen from
\begin{equation}
K = \frac{1}{2M}p^{\mu }p_{\mu } \mbox{\quad} \longrightarrow \mbox{\quad} 
\dot{x}^{\mu } = \dfrac{p^\mu}{M} \ , \ \dot{p}^{\mu }=0 \mbox{\quad}
\longrightarrow \mbox{\quad} \dot x^2 = \text{constant} \ . 
\end{equation}
As discussed by Horwitz, Arshansky, and Elitzur \cite{Two-Aspects}, this framework formalizes a
distinction between two aspects of time:
the time $t$ is one of four spacetime
coordinates $x^\mu$ characterizing the location of a single event, while
%
the time $\tau$ represents the chronological order of multiple events.
%
The physical spacetime event $x^\mu (\tau)$ is understood as an irreversible
occurrence \textbf{at} time $\tau$ so that for $\tau_2 > \tau_1$, event $x^\mu
(\tau_2)$ occurs \textbf{after} $x^\mu (\tau_1)$ and \textbf{cannot change} it.
This changes the significance of a closed timelike curve, resolving the
grandfather paradox.  
The proverbial traveler revisiting at $\tau_3 > \tau_2$ the spacetime locations
$x^\mu (\tau)$ of his grandfather's trajectory as it evolved from $\tau_1$ to
$\tau_2 > \tau_1$, may add events $x^\mu (\tau_3)$ but cannot alter that life trajectory as it
has irreversibly occurred. 
More generally, the 4D block universe $\mathcal{M}(\tau)$ occurs at
$\tau$, evolving to the infinitesimally close the 4D block universe $\mathcal{M}(\tau + d\tau)$
under motion generated by the Hamiltonian $K$.
Because $K$ is a Lorentz scalar and $\tau$ is external, we expect no conflict with general diffeomorphism
invariance.

\section{Classical Off-Shell Electrodynamics}

The origin of the scalar potential $V$ in (\ref{Lagrangian}) can be understood
by requiring invariance under $\tau$-dependent gauge transformations \cite{saad}, leading to
a theory with five gauge fields, $A_\mu(x) \rightarrow a_\mu(x,\tau)$ and 
$a_5(x,\tau)$. 
The maximal gauge freedom of the classical action
\cite{sym12101721}
\begin{equation}
\int d\tau \hspace{1pt} \hspace{1pt} L \ \longrightarrow \ \int d\tau 
\left[L + \frac{d}{d\tau} \Lambda \left( x,\tau \right)\right] = \int d\tau 
\left[L +  \dot x^\mu \partial_\mu \Lambda + \partial_\tau \Lambda \right]
%
%
\end{equation}
suggests a coupling with a pure gauge field with components $\partial_\mu
\Lambda$ and $\partial_\tau \Lambda$.
Introducing the notation
\begin{equation}
\mu ,\nu =0,1,2,3\qquad \text{and} \qquad \alpha ,\beta =0,1,2,3,5
%
%
\end{equation}
and writing $x^5 = c_5 \tau$ in analogy to $x^0 = ct$, we rewrite the classical
interaction as
\begin{equation}
\frac{e}{c}\dot{x}^{\mu }A_{\mu }\left( x\right) - V(x) \ \longrightarrow \ 
\frac{e}{c}\dot{x}^{\mu }a_{\mu }\left( x,\tau \right) +\frac{e}{c}\dot{x}%
^{5 }a_{5 }\left( x,\tau \right) = \frac{e}{c}\dot{x}^{\alpha }a_{\alpha
}\left( x,\tau \right)
\label{cl_int}
\end{equation}
now invariant under 5D gauge transformations
$a_{\alpha } \rightarrow a_{\alpha } +\partial_\alpha \Lambda \left( x,\tau \right)$.
The Lagrangian
\begin{equation}
L = \frac{1}{2}M\dot{x}^{\mu }\dot{x}_{\mu }+ \frac{e}{c}\dot{x}^{\alpha
}a_{\alpha }\left( x,\tau \right)
\label{pM_Lagr}
\end{equation}
admits the conserved 5-current 
\begin{equation}
j^\alpha \left( x,\tau\right) = c \dot{x}^{\alpha } \delta^4 \left( x -
X\left( \tau\right) \right) \qquad \partial_\alpha j^\alpha = \partial_\mu
j^\mu + \partial_5 j^5 = 0
%
%
\end{equation}
which can be related to the Maxwell current by observing that under appropriate
boundary conditions
\begin{equation}
J^\mu (x) = \int d\tau \hspace{1pt} j^\mu \left(
x,\tau\right) \quad \longrightarrow \quad \partial_\mu J^\mu = 0 \ .
\label{cur_cat}
\end{equation}
This integral is called concatenation, understood as the sum of contributions $ g(x,\tau)$ to
$G(x)$ along the worldline, under $g (x,\pm\infty) =0$.
The interaction (\ref{cl_int}) appears to be 5D symmetric,
but this symmetry is broken to vector and scalar representations of O(3,1),
because $\dot x^5 = c_5 \ll c$ is a
constant and not a dynamical quantity.
The Lorentz force \cite{lorentz} found from the Euler-Lagrange equations are
\begin{eqnarray}
M\ddot{x}_{\mu } =&&\hspace{-18pt} \frac{e}{c}\dot{x}^{\beta }f_{\mu \beta }
= \frac{e}{c}\left( \dot{x}^{\nu }f_{\mu \nu }-c_{5}f_{5 \mu}\right) 
\rule[-12pt]{0pt}{12pt}   
\label{L-1} \\
\frac{d}{d\tau }\left( - \frac{1}{2}M\dot{x}^{\mu }\dot{x}_{\mu }\right) =&&%
\hspace{-18pt} c_{5}\frac{e}{c}\dot{x}^{\mu }f_{5 \mu }   
\label{L-2}
\end{eqnarray}
with 5D field strength
\begin{equation}
f_{\alpha\beta} = \partial_\alpha a_\beta - \partial_\beta a_\alpha \qquad
\qquad \alpha, \beta = 0,1,2,3,5 \ .
%
%
\end{equation}
From (\ref{L-2}) we see that $ \varepsilon^\mu \left( x,\tau \right) = f^{5\mu} \left( x,\tau \right)=
\partial^5 a^\mu - \partial^\mu a^5 $ induces mass exchange.  
The field $f_{\mu\nu} \left( x,\tau \right)$ becomes the Maxwell field
$F_{\mu\nu} \left( x \right)$ under concatenation, decoupling from $f^{5\mu}$.

Expanding the interaction term
\begin{equation}
\dot{X}^{\alpha } a_{\alpha } \longrightarrow \int d^{4}x \; \dot{X}%
^\alpha(\tau) \delta^4\left(x-X(\tau)\rule[-4pt]{0pt}{4pt} \right) a_{\alpha
}(x,\tau )
%
%
\end{equation}
we define the sharp current, a delta function on 4D spacetime,
\begin{equation}
j^\alpha (x,\tau) = c \dot{X} ^\alpha(\tau) \delta^4\left(x-X(\tau)\rule[-4pt]{0pt}{4pt} \right)
\label{cur_sharp}
\end{equation}
which by (\ref{cur_cat}) recovers the standard Maxwell current
\begin{equation}
J^\mu (x) = c \int d\tau \p \dot{X} ^\mu(\tau)
\delta^4\left(x-X(\tau)\rule[-4pt]{0pt}{4pt} \right) \ .
%
%
\end{equation}
To complete the electromagnetic action, we introduce a kinetic term for the
electromagnetic field 
\begin{eqnarray}
S_\text{em} \eq \int d^{4}xd\tau \left\{\strt{10}\dfrac{e}{c^2} j^{\alpha }(x,\tau
)a_{\alpha }(x,\tau ) \right. \notag \\
&& \qquad \qquad - \left. \int \frac{ds}{\lambda} \frac{1}{4c} \left[f^{\alpha
\beta }(x,\tau )\Phi (\tau -s)f_{\alpha \beta }\left( x,s\right) \right]
\right\}
\label{action-em}
\end{eqnarray}
where the interaction kernel 
\begin{equation}
\Phi (\tau ) = \delta \left( \tau \right) -(\xi\lambda) ^{2}\delta ^{\prime
\prime }\left( \tau \right)
\qquad \qquad \xi =\dfrac{1}{2}\left[ 1+\left( \dfrac{c_5}{c}\right) ^{2}\right]
%
%
\end{equation}
smooths the sharp current (\ref{cur_sharp}).
The constant $\lambda $ has dimensions of time and serves as a correlation
length along the worldline.
The scalars $j^{\alpha }a_{\alpha }$ and $f^{\alpha \beta }f_{\alpha \beta}$
suggest a 5D symmetry containing O(3,1), either O(4,1) or O(3,2).
Although any higher symmetry is broken by $\dot x^5 =$ constant and by the $\delta^{\prime \prime }\left( \tau
\right)$ term in $\Phi (\tau ) $, it is convenient to introduce a formal 5D
metric
\begin{equation}
g_{\alpha \beta }\underset{\text{flat}}{\longrightarrow } \eta_{\alpha
\beta } = \text{diag}(-1,1,1,1,\sigma) \qquad \qquad \eta_{55} = \sigma =
\pm 1
%
%
\end{equation}
for raising the 5-index in $f^{\alpha \beta }$.
Since $f^{\alpha \beta }f_{\alpha \beta } = f^{\mu \nu }f_{\mu \nu } + 2 \eta^{55}
f_5^{~~\mu }f_{5 \mu }$
we may regard $\sigma = \eta^{55}$ as simply the relative sign of the vector-vector kinetic
term, with no geometric significance.
The interaction kernel is invertible as
\begin{equation}
\varphi (\tau ) = \lambda \Phi^{-1} (\tau ) = \lambda \int \frac{d\kappa}{
2\pi} \,\frac{e^{-i\kappa \tau }}{ 1+\left(\xi \lambda \kappa \right) ^{2}}
= \frac{1}{2\xi }e^{-\vert\tau \vert/\xi \lambda } 
%
%
\end{equation}
\begin{equation}
\int \frac{ds}{\lambda} ~\varphi \left( \tau - s \right) \Phi \left( s
\right) = \delta (\tau) \qquad \int \frac{d\tau}{\lambda} ~\varphi \left(
\tau \right) = 1
%
%
\end{equation}
so that variation of the electromagnetic action with respect to $a_\alpha(x,\tau)$
provides the field equations
\begin{eqnarray}
&&\partial _{\beta }f^{\alpha \beta }\left( x,\tau \right) = \frac{e}{c}\int
ds~\varphi \left( \tau -s\right) j^{\alpha }\left( x,s\right) = \dfrac{e}{c}
\, j_\varphi^{\alpha } \left( x,\tau \right) \rule[-16pt]{0pt}{16pt} 
\label{gauss} \\
&&\partial _{\alpha }f_{\beta \gamma } + \partial _{\gamma }f_{\alpha \beta
} + \partial _{\beta }f_{\gamma \alpha } = 0 \qquad \qquad \text{
(identically)}   
\label{pm-h}
\end{eqnarray}
with source current 
$j_\varphi^{\alpha } \left( x,\tau \right)$ smoothed along
the worldline by convolution of $j^{\alpha } \left( x,\tau \right)$ with the
inverse interaction kernel $\varphi$.  
These are known as pre-Maxwell equations, and when written in 4+1 (spacetime +
$\tau$) components
\begin{equation}
\begin{array}{lcl}
\partial _{\nu }f^{\mu \nu }- \dfrac{1}{c_5} \dfrac{\partial}{\partial \tau}
f^{5\mu }=\dfrac{e}{c} \; j^{\mu }_\varphi & \mbox{\qquad} & \partial _{\mu
}f^{5\mu }=\dfrac{e}{c}\; j^5_\varphi \vspace{8pt} \\ 
\partial _{\mu }f_{\nu \rho }+\partial _{\nu }f_{\rho \mu }+\partial _{\rho
}f_{\mu \nu }=0 &  & \partial _{\nu }f_{5\mu }-\partial _{\mu }f_{5\nu } + 
\dfrac{1}{c_5} \dfrac{\partial}{\partial \tau} f_{\mu \nu }=0%
\end{array}
%
%
\end{equation}
are comparable to Maxwell's equations in 3+1 (space + time) components
where $f_{5\mu} $ plays the role of the electric field sourced by $j^5_\varphi$
in Gauss's
law, and $f^{\mu \nu } $ is a magnetic field induced by ``curl'' and $\tau$
dependence of $f_{5\mu } $.
Writing 
\begin{equation}
f_\Phi^{\alpha \beta }(x,\tau ) = \int \frac{ds}{\lambda} \Phi (\tau
-s)f^{\alpha \beta }\left( x,s\right)
\label{1234}
\end{equation}
translation invariance of the action leads to the Noether symmetry
\begin{equation}
\partial _{\alpha }T_\Phi^{\alpha \beta }=-\frac{e}{c^2}f^{\beta \alpha
}j_{\alpha }
\qquad \qquad 
T_\Phi^{\alpha \beta }= \frac{1}{c}\left(f_\Phi^{\alpha \gamma } f_{~\gamma
}^{\beta }- \frac{1}{4}g^{\alpha \beta }f_\Phi^{\delta \gamma }f_{\delta
\gamma }\right)
\label{noether}
\end{equation}
where $T_\Phi^{\mu \nu }$ is the energy-momentum tensor, and the terms
$T_\Phi^{5 \alpha  }$ represent mass density in the field and the flow of mass
into spacetime.  
Inserting the current (\ref{cur_sharp}) into (\ref{noether}) and using the
second Lorentz force equation (\ref{L-2}), we find
\begin{equation}
\dfrac{d}{d\tau }\left( \int d^{4}x~T^{5\mu }+M\dot{x}^{\mu }\right) =0 %
\mbox{\qquad \ }\dfrac{d}{d\tau }\left( \int d^{4}x~T^{55}-\sigma \hspace{1pt%
} \dfrac{1}{2}M\dot{x} ^{2}\right) =0
%
%
\end{equation}
demonstrating that the total energy-momentum and mass of the particle and field
are conserved \cite{lorentz}.
As was seen for the current, concatenation of the pre-Maxwell equations leads to
the Maxwell equations 
\begin{equation}
\left. 
\begin{array}{c}
\partial _{\beta }f^{\alpha \beta }\left( x,\tau \right) =\dfrac{e}{c}
j_{\varphi}^{\alpha }\left( x,\tau \right) \\ 
\\ 
\partial _{\lbrack \alpha }f_{\beta \gamma ]}=0 \\ 
\end{array}
\right\} \underset{\mathop{\displaystyle \int} \dfrac{d\tau}{\lambda} }{ %
\mbox{\ }\xrightarrow{\hspace*{1.8cm}} \mbox{\ }}\left\{ 
\begin{array}{c}
\partial _{\nu }F^{\mu \nu }\left( x\right) =\dfrac{e}{c}J^{\mu }\left(
x\right) \\ 
\\ 
\partial _{\lbrack \mu }F_{\nu \rho ]}=0 \\ 
\end{array}
\right.
%
%
\end{equation}
representing a sum of microscopic contributions at each
$\tau$ to the Maxwell fields at a given spacetime point.
 
The pre-Maxwell equations lead to the 5D wave equation
\begin{equation}
\partial _{\beta }\partial ^{\beta }a^{\alpha }=(\partial _{\mu }\partial
^{\mu }+\partial _{\tau }\partial ^{\tau })a^{\alpha }=(\partial _{\mu
}\partial ^{\mu } + \frac{\sigma}{c_5^{2}} \; \partial _{\tau
}^{2})a^{\alpha }=- \frac{e}{c}\ j_{\varphi }^{\alpha }\left( x,\tau \right)
%
%
\end{equation}
with Green's function \cite{green}
\begin{eqnarray}
G_{P}(x,\tau ) =&&\hspace{-18pt} -{\frac{1}{{2\pi }}}\delta (x^{2})\delta
(\tau )-{\frac{c_5}{{\ 2\pi ^{2}}}}{\frac{\partial }{{\partial {x^{2}}}}}{%
\theta (-\sigma x^{\alpha }x_{\alpha })}{\frac{1}{\sqrt{%
-\sigma x^{\alpha }x_{\alpha }}}}   \label{greens} \\
=&&\hspace{-18pt} \ G_{Maxwell} + G_{Correlation} 
\end{eqnarray}
where $G_{correlation}$ is smaller than $G_{Maxwell}$ by $c_5 / c$ and drops off
faster with distance.
Notice that $G_{Correlation}$ has spacelike support for $\sigma = -1$ and
timelike support for $\sigma = +1$.
Under concatenation $G_{Maxwell}(x,\tau )$ goes over to the standard Maxwell Green's
function and $G_{Correlation}$ vanishes.

A ``static'' source event evolving along the $x^0$-axis in its rest frame as
$ X\left( \tau \right) =\left(c \tau ,0,0,0\right)$
induces for an observer on the parallel trajectory 
$ x(\tau )=(c\tau ,\mathbf{x})$
a Yukawa-type potential \cite{larry}
\begin{equation}
a^{0}(x,\tau
) ={\dfrac{e}{{4\pi \vert \mathbf{x} \vert }}}\dfrac{1}{2\xi }
e^{-\left\vert \mathbf{x}\right\vert /\xi \lambda c}
%
%
\end{equation}
with photon mass spectrum $m_\gamma c^2 \sim \hbar / \xi \lambda $.
Using the accepted experimental error for photon mass $ \Delta m_\gamma \sim 10^{-18}$ eV leads
to $\lambda > 10^4$ seconds.
The constant $\lambda$ can be seen as a correlation time along the worldline,
the width of the ensemble of events contributing to the pre-Maxwell current and
potential \cite{entropy}. 
In the limit $\lambda \rightarrow 0$ the kinetic term in the action
(\ref{action-em}) reduces to $f^{\alpha\beta}f_{\alpha\beta}$, the photon mass
spectrum goes to infinity, and $a^0$ becomes a delta function.  
In the limit $\lambda \rightarrow \infty$ the photon mass spectrum vanishes and
the pre-Maxwell system reduces to Maxwell electrodynamics.
The Li\'{e}nard-Wiechert potential for an arbitrary source event $X^{\mu }\left(
\tau \right)$ at an observation point $x^\mu$ similarly leads to the Maxwell
formula multiplied by the factor $\varphi \left( \tau -\tau_R\right) $ where $
\tau_R$ is the retarded time found from $\left[ x - X(\tau_R) \right]^2 = 0$.
Comparing the Lorentz forces for $e^-/e^+$ and $e^-/e^-$ scattering leads to an
experimental bound on $c_5 \ll c$ \cite{speeds}.

\section{Mass interactions and mass stability} 

A simple model of mass variation is a uniformly moving particle undergoing a
stochastic perturbation $x = u\tau \rightarrow u\tau + X(\tau)$ 
when entering a dense distribution of charged particles \cite{mass}.  
If the typical short distance between charge centers is $d$ then the particle
will encounter charges periodically, with a short 
characteristic period $d / \left\vert \mathbf{u}\right\vert $, leading to a high
characteristic frequency $\omega _{0}=2\pi \left\vert \mathbf{u}
\right\vert  / d$.
Expanding the perturbation in a Fourier series
\begin{equation}
X\left( \tau \right) =\text{Re}\sum_{n}a_{n}~e^{in\omega _{0}\tau }
\qquad \qquad
a^\mu_{n} = 
\alpha d\left( s_{n}^{0},\mathbf{s}_{n}\right)
=\alpha d\left( cs_{n}^{t},\mathbf{s}_{n}\right)
\end{equation}
with normalized coefficients $s_{0}^{\mu }\sim 1$ and some macroscopic factor
$\alpha \lesssim 1$. 
We obtain a small perturbation of position $\left\vert X^{\mu }\left( \tau
\right) \right\vert \sim \alpha d$, but a velocity perturbation 
\begin{equation}
\dot{x}^{\mu }\left( \tau \right) = u^{\mu }+\alpha \left\vert \mathbf{u}%
\right\vert ~\text{Re}\sum_{n}2\pi n~s_{n}^{\mu }~ie^{in\omega _{0}\tau }
\end{equation}
of macroscopic scale $\alpha \left\vert \mathbf{u} \right\vert$.
Writing the particle mass as $m(\tau) = - M \dot x ^2 / c^2$ leads to a
macroscopic mass perturbation
\begin{equation}
m\longrightarrow m\left( 1+\frac{\Delta m}{m}\right) \mbox{\qquad}\frac{
\Delta m}{m}=4\pi \alpha \left\vert \mathbf{u}\right\vert \text{Re}
\sum_{n}n~s_{n}^{t}\ ie^{in\omega _{0}\tau } 
\end{equation}
which could persist when the particle leaves the charge distribution.

One possible reason that we do not see such mass perturbations more frequently
is a self-interaction that tends to restore mass to its familiar on-shell value.
We consider a particle with arbitrary $\dot x^0 (\tau)$ in its rest frame, where
$\ddot x^0 \ne 0$ entails mass variation through $ -M\dot x^2 = Mc^2 \p \dot t^2 (\tau)$.
Along the worldline, the particle may interact at time $\tau^\ast > \tau$ with
the field it produces at $\tau$, but of course $G_{Maxwell} $, the leading term in the Green's
function, vanishes on $\Delta X (\tau^\ast, \tau) = X(\tau^\ast) - X(\tau)= c(
t(\tau^\ast) - t(\tau) , \mathbf{0}) $,  the timelike separation.
Nevertheless, from (\ref{greens}) we see that $G_{Correlation}$ has support
for $-\sigma x^{\alpha }x_{\alpha } > 0 $, which is this case is the condition
\begin{equation}
- \big[ \Delta X ^2 + c_5^2 (\tau^\ast -\tau)^2\big] = c^2 \left( \big[ t\left( \tau ^{\ast
}\right) -t\left( \tau\right) \big] ^{2}-\frac{c_5^{2}}{c^{2}}(\tau ^{\ast
}-\tau)^{2}\right) > 0 
\label{rad-cond}
\end{equation}
when $\sigma =+1$.
Expanding $t(\tau)$ in a Taylor series, one finds that condition
(\ref{rad-cond}) is satisfied if and only if $\ddot t \ne 0$, which in the rest
frame implies mass variation. 
Now suppose that a particle evolves uniformly as $t = \tau$ until $\tau = 0 $
when it makes a sudden jump to $t = (1+\beta)\tau$. 
The field strength acting on the particle at $\tau^\ast > 0$ contains only the
component
\begin{equation}
f^{50} \approx \frac{e}{4\pi ^2}\dfrac{1}{c_5^2\left( \beta \tau ^\ast
\right) ^3} \ Q \left( \beta,\dfrac{c_5^2}{c^2} \right)
\end{equation}
where $Q\left( \beta ,\dfrac{c_5^{2}}{c^{2}}\right)$ is a positive function that
vanishes for $\beta = 0$ or $c_5 = 0$.
The Lorentz force is then
\begin{eqnarray}
M\ddot{x}^{0} =&&\hspace{-18pt} -c_5ef^{50}= \left\{ 
\begin{array}{lcc}
0 & , & \tau^\ast < 0\vspace{4pt} \\ 
-\dfrac{\lambda e^2}{4\pi ^2}\dfrac{1}{c_5\left( \beta \tau ^\ast \right) ^3}
\ Q \left( \beta,\dfrac{c_5^2}{c^2} \right) & , & \tau^\ast > 0
\end{array}
\right. \\
M\ddot{x}^{i} =&&\hspace{-18pt} -c_5ef^{5i} \dot x_i = 0 
\end{eqnarray}
and
\begin{equation}
\frac{d}{d\tau }\left( -\frac{1}{2}M\dot{x}^{2}\right) =
ef^{5\mu }\dot{x} _{\mu }=-ecf^{50}\dot{t}=-{\frac{\lambda e^2}{{4\pi ^{2}}}}
\dfrac{c}{c_5^2\left( \beta \tau ^\ast \right) ^3} \ Q \left( \beta,\dfrac{
c_5^2}{c^2} \right) \dot t
%
%
\end{equation}
which acts as a restoring force, damping the mass toward its on-shell value and
vanishing on shell.

Another approach \cite{chemical} describes the particle as a statistical
ensemble with both an equilibrium energy and an equilibrium mass,
controlled by temperatures and chemical potentials, assuring asymptotic
states with the correct mass.
The thermodynamic properties are found from the microcanonical ensemble, where
both energy and mass are parameters of the distribution. 
A critical point in the free energy emerges from equilibrium
requirements of the canonical ensemble (where total system mass is
variable), and equilibrium
requirements of the grand canonical ensemble (where a chemical potential arises
for the particle number).
Because particle mass is
controlled by a chemical potential, asymptotic variations in the mass
are restored to a given value by relaxation, satisfying the equilibrium
conditions.

\section{Off-Shell Quantum Electrodynamics} 

Transforming the classical Lagrangian (\ref{pM_Lagr}) to Hamiltonian form, we
are led to the Stueckelberg-Schrodinger equation
\begin{equation}
\left(i\hbar \partial _{\tau }+e\frac{c_5}{c}a_5 \right)\ \psi \left(x,\tau
\right)=\frac{1}{2M} \left(p^{\mu }-\frac{e}{c}a^{\mu }\right)\left(p_{\mu }-%
\frac{e}{c}a_{\mu }\right)\ \psi \left(x,\tau \right)
%
%
\end{equation}
with local 5D gauge invariance
\begin{equation}
\psi (x,\tau ) \rightarrow e^{ie\Lambda (x,\tau ) / 
\hbar c} \, \psi (x,\tau )
\qquad 
a_{\alpha }(x,\tau ) \rightarrow a_{\alpha
}(x,\tau )+\partial _{\alpha }\Lambda (x,\tau )
%
%
\end{equation}
and global gauge invariance providing the conserved current $\partial _{\alpha
}j^{\alpha } = 0$ with 4-vector part
\begin{equation}
j^{\mu }=-\dfrac{i\hbar}{2M}\left\{\psi ^*\left(\partial ^{\mu }-\dfrac{ie}{c%
}a^{\mu }\right)\psi -\psi \left(\partial ^{\mu }+\dfrac{ie}{c}a^{\mu
}\right)\psi ^*\right\} 
%
%
\end{equation}
and event density $ j^5 =c_5\left\vert\psi (x,\tau)\right\vert^{2}$ representing the
probability of finding an event at a spacetime point $x$ at time $\tau$.  
The quantum Lagrangian is 
\begin{equation}
\mathcal{L}= \psi ^{*}(i\partial _{\tau }+ea_{5})\psi -\frac{1}{2M}\psi
^{*}(-i\partial _{\mu }-ea_{\mu })(-i\partial ^{\mu }-ea^{\mu })\psi -\frac{
\lambda }{4}f^{\alpha \beta } f^\Phi_{\alpha \beta }
%
%
\end{equation}
where $ f^{\Phi }_{\alpha \beta }\left( x,\tau \right)$ is defined in
(\ref{1234}),
which admits Jackiw first order constrained quantization \cite{jackiw} by introducing $\epsilon ^{\mu
}=f^{5\mu } $.  
Because $\dot a^5$ does not appear in the Lagrangian, path integration over $a^5$ inserts the 
Gauss law constraint $ \delta(\partial^{\mu }\epsilon^{\Phi }_{\mu } - e\psi
^{*}\psi )$ which may be solved to eliminate longitudinal modes.  
Feynman rules may be read from the unconstrained Lagrangian 
\begin{equation}
\mathcal{L} =i\psi ^{*}\dot{\psi}-\frac{1}{2M}\psi ^{*}(-i\partial _{\mu
}-ea_{\perp }\,_{\mu })(-i\partial ^{\mu }-ea_{\perp }^{\mu })\psi+\frac{1 }{%
2}a_{\perp }\,_{\mu }\left( \Box +\sigma \partial _{\tau }^{2} \right)
a_{\perp }^{\Phi }\,^{\mu }
%
%
\end{equation}
as matter and photon propagators
\begin{equation}
\dfrac{1}{(2\pi )^{5}}\dfrac{-i}{\frac{1}{2M}p^{2}-P-i\epsilon }
\qquad \qquad 
\left[ g^{\mu \nu }-\dfrac{k^{\mu
}k^{\nu }}{k^{2}}\right] \dfrac{-i}{k^{2}+\kappa ^{2}-i\epsilon }\ \ \dfrac{1%
}{1+\lambda ^{2}\kappa ^{2}}
%
%
\end{equation}
along with three and four particle vertex factors
\begin{equation}
\begin{array}{c}
\dfrac{e}{2M}i(p+p^{\prime })^{\nu }\ (2\pi )^{5}\delta^{4}(p-p^{\prime
}-k)\delta (P-P^{\prime }-\kappa ) \strt{12}\\
\dfrac{-ie^{2}}{M}(2\pi )^{5}g_{\mu \nu }\delta
^{4}(k-k^{\prime }-p^{\prime }+p)\delta (-\kappa +\kappa ^{\prime
}+P^{\prime }-P)
\end{array}
%
%
%
\end{equation}
which conserve total energy-momentum and mass.  
The matter propagator
\begin{equation}
G(x,\tau) = \mathop{\displaystyle \int} \frac{d^4k \hspace{1pt} d\kappa}{%
(2\pi )^{5}} \dfrac{e^{i(k \cdot x - \kappa \tau)}}{\frac{1}{2M}%
k^{2}-\kappa-i\epsilon } =i \theta(\tau)\mathop{\displaystyle \int} \frac{%
d^4k}{(2\pi )^{4}}e^{i(k \cdot x - \frac{1}{2M}k^{2}+i\epsilon)}
%
%
\end{equation}
enforces retarded causality in $\tau$, so that there are no matter loops, just
as there are no grandfather paradoxes.
This expression was previously found by Feynman \cite{Feynman-1} for the
Klein-Gordon equation, leading to the Feynman propagator by extracting a stationary eigenstate
of the mass operator $-i \hbar \partial_\tau$ as
\begin{equation}
\int_{-\infty}^\infty d\tau e^{-i(m^2/2M)\tau} G(x,\tau) = \int \frac{d^4 k}{%
(2\pi)^4} \frac{e^{ik\cdot x}} {\frac{1}{2M} (k^2+ m^2) -i\epsilon} =2M \
\Delta_{\mathrm{F}} (x)  \ . 
%
%
\end{equation}
We see that the interaction kernel inherited from the classical electromagnetic
term provides the natural mass cut-off $\left(1+\lambda
^{2}K^{2} \right)^{-1} $ which renders the theory super-renormalizable.
The cross-section for elastic scattering is nearly identical to the Klein-Gordon
case, but the pole is slightly shifted away from $0^\text{o}$ for non-zero
mass exchange between the outgoing particles (expressed as an undetermined hyperangle,
much as the scattering angle is undetermined in on-shell QED)
\cite{qft}.

\section{General relativity with $\tau$-evolution}

As described in Section \ref{s:SHP}, the SHP framework poses 
a block universe $\mathcal{M}(\tau)$ that evolves to a block universe $\mathcal{M}
(\tau + d\tau)$ under a Hamiltonian $K$.
We thus expect that the spacetime metric $g_{\mu\nu}(x,\tau)$ should similarly evolve
to $g_{\mu\nu}(x,\tau + d\tau)$.
To find field equations that prescribe this evolution, we look for hints from
the development of the pseudo-5D off-shell electromagnetic field equations, which differ
from Maxwell equations written in five dimensions because excluding $x^5$ from
the dynamical degrees of freedom breaks any 5D symmetry \cite{AS,sym12101721}.
Just as there is no Lorentz force for $\ddot x^5$, there must be no geodesic
equation for $\ddot x^5$ in curved spacetime.

In standard 4D general relativity (GR), the invariance of the squared interval
$\ \delta x^2 = \gamma_{\mu\nu} \delta x^\mu \delta x^\nu = \left( x_2 - x_1
\right)^2 $ between neighboring events (an instantaneous displacement) is a
geometrical statement, characterizing the manifold $\mathcal{M}$.
For events $ X_1 = (x_1,c_5 \tau_1) $ and $ X_2 = \left( x_2 , c_5
\left( \tau_1 + \delta \tau\right) \right) $ belonging to $\mathcal{M}(\tau)$ and
$\mathcal{M} (\tau + d\tau)$ the squared interval 
\begin{equation}
dX^\alpha dX_\alpha = \left( \delta x + \dfrac{d{\ x} (\tau)}{d\tau} \delta \tau
\right)^2 + \sigma c_5^2 \delta \tau^2 = g_{\alpha\beta} \left(
x,\tau\right) \delta x^\alpha \delta x^\beta
%
%
\end{equation}
suggests a pseudo-5D metric $g_{\alpha\beta} \left( x,\tau\right) $, analogous
to the pseudo-5D electromagnetic field $f^{\alpha\beta} \left( x,\tau\right) $.
The evolution of $g_{\alpha\beta} \left( x,\tau\right)$ differs from a standard
metric defined in 5D, because it combines 4D geometrical symmetries of
$\mathcal{M}(\tau)$ with the scalar dynamical symmetry of Hamiltonian $K$.    
%
%
To preserve the constraint $ x^5 \equiv c_5 \tau  $ we expand the classical
Lagrangian as
\begin{equation}
L=\dfrac{1}{2}Mg_{\alpha\beta}(x,\tau)\dot{x}^{\alpha}\dot{x}^{\beta} = 
\dfrac{1}{2}Mg_{\mu\nu}\; \dot{x}^{\mu}\dot{x}^{\nu} + Mc_5 \; g_{\mu 5}\dot{%
x}^{\mu} + \dfrac{1}{2}Mc^2_5 \; g_{55}
%
%
\end{equation}
to obtain four geodesic equations and an identity
\begin{equation}
0 
=\ddot{x}^{\alpha}+\Gamma
_{\beta\gamma}^{\alpha}\dot{x}^{\beta}\dot{x} ^{\gamma} \longrightarrow
\left\{ 
\begin{array}{l}
\ddot{x}^{\mu}+\Gamma _{\lambda \sigma }^{\mu }\dot x^\lambda \dot x^\sigma
+2c_5\Gamma _{5\sigma }^{\mu }\dot x^\sigma +c^2_5\Gamma _{55}^{\mu } = 0 
\rule[-12pt]{0pt}{12pt} \\ 
\ddot{x}^5 = \dot c_5 \equiv 0 
\end{array}
\right. 
%
%
\end{equation}
and the Hamiltonian
\begin{equation}
K = p_{\mu }\dot{x}^{\mu }-L = \dfrac{1}{2}M g_{\mu \nu }\dot x^{\mu }\dot
x^{\nu } - \dfrac{1}{2}Mc^2_5 \; g_{55} = L - Mc^2_5 \; g_{55}
%
%
\end{equation}
from which we find
\begin{equation}
\frac{d K}{d \tau } = -\frac{1}{2}M\dot{x}^{\mu }\dot{x}^{\nu }\frac{
\partial g_{\mu \nu }}{\partial \tau }-\frac{1}{2}Mc_{5}^{2}\frac{\partial
g_{55}}{ \partial \tau }
%
%
\end{equation}
showing that particle mass is not generally conserved along geodesics.

We define the event density in spacetime $n(x,\tau)$ and mass density $\rho(x,\tau) =M
n(x,\tau)$,  
leading to the event current $ j^{\alpha }\left( x,\tau
\right) =\dot{x}^{\alpha }(\tau) \rho(x,\tau) $ and continuity equation 
$ {\boldsymbol{\nabla}}_{\alpha }j^{\alpha } = 0 $
with covariant derivative ${\boldsymbol{\nabla}}_{\mu }$ defined in the standard
manner and ${\boldsymbol{\nabla}}_{5} = \partial_5$.
Current conservation leads to conservation of the analogously defined
mass-energy-momentum tensor $T^{\alpha \beta }=\rho \dot{x}^{\alpha }
\dot{x}^{\beta }$.
%
If we write the standard Einstein field equations in 5D and study the linearized
form for weak gravitation $g_{\alpha\beta} \approx \eta_{\alpha\beta} + h_{\alpha\beta}$,
we obtain a wave equation that can be solved using the Green's function (\ref{greens}).
However, the metric perturbation found from a ``static'' source in its rest
frame includes $h_{00} = 2h_{ij} = h_{55}$, which deviates from the expected structure,
$h_{00} = h_{ij} \gg h_{55}$.
To determine the correct modification of the field equations we choose a form
that preserves the 5D symmetries of the Ricci tensor $R_{\alpha\beta} $, but breaks the apparent 5D
symmetry in the relationship between $R_{\alpha\beta} $ and $T_{\alpha\beta} $.
In trace-reversed form, we write \cite{sym12101721}
\begin{equation}
R_{\alpha\beta} = \frac{8\pi G}{c^4} \left( T_{\alpha\beta} - \frac{1}{2}
\bar g_{\alpha\beta}\bar T\right)
\qquad \bar g_{\mu \nu} =  g_{\mu \nu} \qquad \bar g_{5 \alpha } = 0
\label{modified}
\end{equation}
where $ \bar T = \bar g^{\mu \nu}T_{\mu \nu} $, which for the ``static''
source in the weak field approximation leads to
\begin{equation}
g_{\mu\nu} = \left( -\left( 1-\frac{2Gm}{c^2r}\right) ,\left( 1-%
\frac{2Gm}{c^2r}\right)^{-1}\delta_{ij} \right)
\qquad 
g_{55} = \sigma + o\left( \frac{c^2_5}{c^2}\right)
%
%
\end{equation}
consistent with a spherically symmetric Schwarzschild metric.
A source in its rest frame with mass varying arbitrarily as $\dot x^0(\tau) =
c[1+\alpha \left( \tau \right) / 2] $ leads to a
$\tau$-dependent perturbation.  The geodesic equations for a test particle in
this space undergoes a nonlinear $x^0$ acceleration, and satisfies a radial equation
\begin{equation}
\frac{d}{d\tau }\left\{ \frac{1}{2}\dot{R}^{2}+\frac{1}{2}\frac{L^{2}}{
M^{2}R^{2}}-\frac{GM}{R}\left( 1+\frac{1}{2}\alpha \left( \tau \right)
\right) \right\} = -\frac{GM}{2R}\frac{d}{d\tau }\alpha \left( \tau \right)
%
%
\end{equation}
with conserved angular momentum $ L = MR^2 \dot \phi$.
The term in brackets on the LHS is the Hamiltonian in these coordinates,
indicating that the mass of the test particle is not conserved when the mass of
the source varies.
This simple example suggests that a source particle of varying mass can transfer mass
across spacetime to a test mass moving geodesically under the influence of the
metric field induced by the source \cite{AS}. 

Decomposing these field equation (\ref{modified}) into 4+1 form
\cite{sym12101721}, analogous to the 3+1
decomposition used in the ADM formalism \cite{ADM}, we find that the 20 spacetime
components $ R_{\mu\nu}$
are unconstrained second order evolution equations, while the five components
$ R_{\alpha 5}$
are constraints that propagate at first order in $\partial_\tau$.  
Moreover, from $\bar T = T - g^{55} T_{55}$, 
the mass density $T_{55}$ sourced by $g_{5 5}$ and not necessarily constant,
is seen to play the role of a small cosmological term $\Lambda$.



\end{document}